\documentclass[aps,prc,showpacs,superscriptaddress,preprint,10pt]{revtex4-1}
\usepackage{graphicx}
\usepackage{dcolumn}
\usepackage{bm}
\usepackage{color}
\usepackage{comment}
\usepackage{ulem}
\usepackage{here}
\usepackage{bm}
\usepackage{braket}
\usepackage{graphicx}

\begin{document}

\title{Dependence of total kinetic energy of fission fragments on
the excitation energy of fissioning systems
}

\author{Kazuya Shimada}
\affiliation{
Nuclear Engineering Course, Transdisciplinary Science and Engineering, School of Environment and Society, Tokyo Institute of Technology, Tokyo, 152-8550 Japan
}
\author{Chikako Ishizuka}
\email{ishizuka.c.aa@m.titech.ac.jp}
\affiliation{
Institute of Innovative Research, Tokyo Institute of Technology, Tokyo, 152-8550 Japan
}
\author{Fedir A. Ivanyuk}
\affiliation{
Institute for Nuclear Research, 03028 Kiev, Ukraine}
\author{Satoshi Chiba}
\affiliation{
Institute of Innovative Research, Tokyo Institute of Technology, Tokyo, 152-8550 Japan
}

\date{\today}

\begin{abstract}%
We elucidated the reason why the average total kinetic energy (TKE) of  fission fragments decreases when the excitation energy of the 
fissioning systems increase as indicated by experimental data for the neutron-induced fission events. 
To explore this problem, we used a method based on the four-dimensional Langevin equations we have developed.
We have calculated the TKE of fission fragments for fissioning systems $^{236}$U$^{*}$
 and $^{240}$Pu$^{*}$ excited above respective fission barriers, and compared the results with experimental data for n + $^{235}$U and n + $^{239}$Pu reactions, respectively.
From the Langevin-model analysis, 
we have found that the shape of the abundant heavy fragments changes from almost spherical for low excitation domain to highly prolate shape for high excitation energy, 
while that of the light fragments does not change noticeably. 
The change of the "shape" of the heavy fragments causes an increase of a distance between the charge centers of the nascent fragments
just after scission as excitation energy increases. 
Accordingly, the Coulomb repulsion between the two fragments decreases with an increase of the excitation energy, 
which causes the decrease of the average TKE. 
In this manner, we found that the change of the shape of the heavy fragment as a function of the excitation energy is the key issue for the TKE of fission fragments to decrease as the excitation energy of the fissioning nuclei increases.  
In other words, washing out of the shell effects which affect the shape of the heavy fragments is the key reason for the decreasing energy dependence of the average TKE of the fission fragments.
\end{abstract}


\maketitle

\section{Introduction}
Nuclear fission is the fundamental process both in nuclear energy and fission recycling during the r-process nucleosynthesis. 
Therefore, it is necessary to understand the nuclear fission process in detail to utilize it with a good accuracy in various fields.
In their earlier works just after the discovery of nuclear fission, Bohr, Wheeler, and Hill~\cite{Bohr-Wheeler,Hill-Wheeler} proposed the basic concepts of nuclear fission such as saddle point, fission barriers, fissility, and zero-point oscillation of shape parameters based on the liquid-drop model. 
They found that each mode may have independent zero-point oscillation energy~\cite{Hill-Wheeler}.
These ideas are
still important even today.
However, precise understanding of the nuclear fission process is still challenging because of its complexity as a large-amplitude collective motion of quantum  systems consisting of a finite number of nucleons~\cite{Bender}. 
%
%
%

It is a special feature of nuclear fission that a large amount of energy is released as the Q-value and that most of this energy turns into the form of total kinetic energy (TKE) of the fission fragments. 
According to this, the sum of the Q value of fission and the initial excitation energy $E^*$ is divided into TKE and other energies. 
In particular, $Q + E^* - \rm{TKE}$ is the total excitation energy of fission fragments, and this excitation energy causes the emission of prompt neutrons and $\gamma$ rays~\cite{Okumura}.
The number of prompt neutrons is important for the criticality of nuclear reactors, and $\gamma$ rays are also one of the heat sources of  nuclear reactors. 
$\beta$-decay after gamma emission is also a phenomenon that occurs according to the distribution of nuclides formed by the emission of prompt neutrons.
Therefore, a quantitative understanding of TKE is important to elucidate the nature of the fission process.

There remain unsolved questions in the study related to the TKE of the fission process.
One of them is the relationship between the excitation energy of fissioning nuclei and the TKE.
Intuitively, it is expected that the TKE increases as the excitation energy of the fission nucleus increases since the energy available to any mode increases.  
In the terminology of the 4-dimensional Langevin theory which we use, $Q+E^*$ 
is shared among 4 collective coordinates and intrinsic energy, one of the collective
coordinate being the elongation of the whole system.  Therefore, if the total energy
available increases, there is more chance for the elongation degree-of-freedom to get more energy so that the translational motion, namely, kinetic energy of fragments, gets more energy.
Therefore, it is a big mystery to know that the TKE decreases in neutron-induced fission events as the incident neutron energy increases which lead to an increase of the excitation energy of the compound nuclei ~\cite{Dyachenko,Surin,Akimov,Duke235U+n,Duke238U+n,Duke239Pu+n}. 
This fact contradicts the above intuitive expectation.  
Of course, nuclear fission is a very complex motion of finite number of nucleons, so such a naive picture may be an oversimplification of the phenomenon.
As a matter of fact, it is expected that friction will increase as the excitation energy increases, leading to the decrease of pre-scission kinetic energy.
Nevertheless, the effect of friction on TKE reduction is a minor one since the main source of TKE is Coulomb repulsion between the nascent fragments formed just after scission. 
%

Various approaches have been proposed to study the fission process and properties of fission fragments such as mass and TKE distributions. 
For a quantitative estimation of TKE, we need a dynamical model that can describe a process starting from an almost spherical compound nucleus to scission via fission saddle points.
For example, Langevin equations based on the fluctuation-dissipation theorem have been successfully applied for this purpose ~\cite{Usang2016,Usang2017,Ishizuka2017,Usang2019,Ishizuka2020}.    
%
They have theoretically studied nuclear fission using a three-dimensional Langevin model with microscopic transport coefficients~\cite{Usang2016,Usang2017} and a four-dimensional Langevin model with macroscopic transport coefficients~\cite{Ishizuka2017}.
Our previous studies found that both systematic and anomalous trends in fission fragment mass distribution and TKE
could be understood by a correlated twin transition~\cite{Usang2019}.
In the same context, we predicted a new super asymmetric fission mode in a region of superheavy nuclei where the magicity of $^{208}$Pb plays an important role~\cite{Ishizuka2020}.
In the present study, we apply the same methodology for typical actinide nuclei important for energy applications to elucidate the relationship between the excitation energy of a compound nucleus and the TKE of the fission fragments focusing on the ``shape'' of the fission fragments.
%
%

This paper is organized as follows.
Sec. II., the computational method based on  the four-dimensional Langevin model used in this study is explained. 
In Sec. III, the calculation results are shown and compared with the experimental data.
Especially, we focus on the deformation of fission fragments and discuss the relationship between the change of the fragment deformations on excitation energy and its impact on the TKE.
In Sec. IV, the conclusions and future prospects of this study are stated.

\section{Computational Method}
To describe the nuclear shape during fission process, we adopt a shape parametrization of the so-called two-center shell model~\cite{Maruhn}, which is basically two oscillators connected smoothly to each other in the region between them.
In this parametrization, the nuclear shape is described by 5 parameters, namely, the elongation $z_0$,
deformation of the outer tips of nuclei, $\delta_1$ and $\delta_2$, mass asymmetry $\alpha$ which is defined as $(A_1-A_2)/(A_1+A_2)$ where $A_1$ and $A_2$ are mass numbers of the two parts of the fissioning system which eventually leads to two nascent fragments, and the neck parameter $\epsilon$. 
The parameter $z_0$ corresponds to the distance between centers of the two oscillators normalized by $R_0 = 1.2 \cdot A_c^{1/3}$, where $A_c=A_1 + A_2$ 
stands for mass number of the fissioning nucleus.  See ref. \cite{Ishizuka2017} for details.
Out of these 5 parameters, we fixed $\epsilon=0.35$, and the other 4 parameters were selected as the dynamical variables whose time evolutions are described by the Langevin equation.
The value of $\epsilon$=0.35 was determined from our previous analysis for the mass distributions of actinide nuclei.  
The four dynamical parameters may be denoted in general as $\{q_{\mu}:~\mu=1, \cdots, 4\}=\{z_0, \delta_1, \delta_2, \alpha\}$, then, the Langevin equations take the following form:
\begin{widetext}
\begin{eqnarray}
\frac{dq_\mu}{dt} 
&=& (m^{-1})_{\mu\nu}p_\nu , \\
\frac{dp_\mu}{dt}
&=& -\frac{\partial F(q,T)}{\partial q_\mu}
- \frac{1}{2}\frac{\partial (m^{-1})_{\nu\sigma}}{\partial q_\mu}p_\nu p_\sigma \nonumber \\
& & -\gamma_{\mu\nu}(m^{-1})_{\nu\sigma} p_\sigma
+g_{\mu\nu}R_\nu(t), 
\end{eqnarray}
\end{widetext}
where all Greek letters take values of 1 to 4, and summation for repeated subscripts is implicitly assumed. 
We solve the Langevin equations using the collective inertia tensor $m_{\mu\nu}$ calculated within the Werner-Wheeler approximation~\cite{Werner-Wheeler1,Werner-Wheeler2}
and the friction tensor $\gamma_{\mu\nu}$ based on the wall-and-window formula~\cite{wall-and-window}. 
The symbol $R_\nu$ stands for white noise, while $g_{\mu \nu}$ denotes its strength. 
The dumping factor $k_s$ for the wall-and-window friction was determined to be 0.27~\cite{ffric1} from analysis of the widths of giant resonances by Nix and Sierk, and we have used this value in our past researches. 
They solved the Fokker-Plank equation with a shape parametrization different from the two center model used here.
The value of this dumping factor should depend on the combination of other parameters, and also on the parametrization and models to be solved.  
In this paper, we adjusted this parameter to 0.55 to reproduce the average TKE value for n + $^{235}$U and n + $^{239}$Pu reaction at neutron energy region below 6 MeV.  
The change of $k_s$ from 0.27 to 0.55 gives a constant shift to average TKE by about 3 to 4 MeV, but does not affect the dependence of the TKE on excitation energy.
Therefore it does not affect the purpose of this work. 
We may use a different value of the $k_s$ parameter in future work when different values of other parameters such as $\epsilon$ are selected to reproduce other observables.
Anyway, this work aims to understand the dependence of the average TKE of fission fragments on excitation energy.
It is not affected by choice of $k_s$ as long as we keep it to be a constant value.

We introduced independent effective temperature for each dynamical variable in the following manner.
The strength of the random force term $g_{\mu \nu}$ is related to the friction tensor $\gamma_{\mu \nu}$ via a combination of the Einstein relation and fluctuation-dissipation theorem:
\begin{equation}
g_{\mu \sigma} g_{\sigma \nu} = T \gamma_{\mu \nu}    
\label{eq:es}
\end{equation}
where $T$ denotes the thermodynamic temperature or, in other words, intrinsic temperature of the heat bath. 
From this equation, it is known that the strength $g_{\mu \nu}$ depends on the temperature in a manner proportional to $\sqrt{T}$.
This $T$ is derived from the initial excitation energy $E_x$ as
\begin{eqnarray}
E_{int} &=& E_x - \frac{1}{2}\left(m^{-1}\right)_{\mu\nu}p_\mu
p_\nu - F(q,T=0) \nonumber \\
&=& aT^2  \ .
\end{eqnarray}
Here, $F$ and  $a$ denote the free energy of the system at $T$=0 and the level density parameter, respectively.
Then, in order to introduce "effective" temperature which accounts for quantum correction to each dynamical variable corresponding to zero-point oscillation, we first solve the above equation (\ref{eq:es}) for $T=1$ MeV and obtain a "pseudo" strength $g'$, namely,
\begin{equation}
g'_{\mu \sigma} g'_{\sigma \nu} = \gamma_{\mu \nu}    
\end{equation}
Nextly, we introduce effective temperature $T^{eff}_{\mu}$\cite{ffric2,ffric3} for each random variable $q_\mu$ by using the zero-point energy $\frac{1}{2} \hbar \omega_\mu$ for each variable as  
\begin{equation}
T^{eff}_\mu = 
\frac{1}{2}\hbar\omega_\mu\coth{\frac{\hbar\omega_\mu}{2T}}.
\end{equation}
Then, the strength $g$ is obtained as $g_{\mu \nu}=\sqrt{T^{eff}_{\mu}} g'_{\mu \nu}$ (do not take sum on $\mu$).
In this paper, we have investigated the reasonable zero-point energy $\hbar\omega_\mu/2$ of oscillators forming the heat bath, for each degree of freedom. 
Our previous works~\cite{Ishizuka2017} estimated the zero-point energy as 1 MeV taking the middle of 0.45 to 2.23 MeV suggested in ref.~\cite{Bohr-Wheeler,Hill-Wheeler}.  
As is already discussed in their pioneering works, 
the magnitude of the zero-point energy can be different for different degrees-of-freedom, since it accounts for the curvature of the potential energy landscape for each of the independent dynamical variables.
Based on the previous study, we compare two sets of $\{\hbar\omega_\mu \}= (2, 2, 2, 2)$ MeV and $(1, 1, 1, 2)$ MeV. 
We found that the width of the mass distribution is somehow sensitive to $\hbar\omega_4$.
For example, the distribution becomes too sharp for $\hbar\omega_4=1$ MeV, and the symmetric components of the distribution become too large for $\hbar\omega_4=3$ MeV, while $\hbar\omega_4=2$ MeV can reproduce both the width and peak structure reasonably. 
Then, we decided to fix $\hbar\omega_4=2$ MeV.
The other parameters $\hbar\omega_i$ ($i = 1, 2, 3$) do not affect much important observables except for a constant shift in the TKE, which was taken care of by the $k_s$ parameter as described above.
We take a set of $\{\hbar\omega_\mu \} = (1,1,1,2)$ [MeV] for the calculations below.
\section{Results and discussion}
We compare the fission fragment mass distributions (FFMDs) and the dependence of the averaged total kinetic energies of fission fragments on the incident neutron energy $E_n$, calculated by the present model and experimental data. 
In this comparison, we shift the calculated data or experiment by the neutron separation energy since our calculations were 
performed for $^{236}$U and $^{240}$Pu excited above their fission barriers as compound nuclei, so they are denoted as 
$^{236}$U$^*$ and $^{240}$Pu$^*$.
In the following, we label the fissioning system by the compound nuclei, while experimental data were taken from neutron-induced reactions.

\begin{figure}[htbp]
\centering\includegraphics[scale=0.75,angle=0]{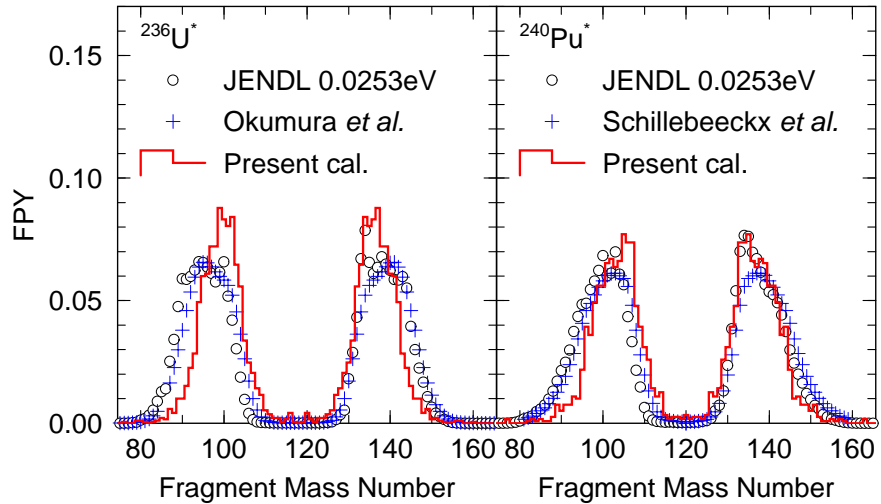}
%
\caption{(Color online) The mass distribution of fission fragments originated from the compound nuclei $^{236} {\rm{U}}$$^*$ (left) and $^{240} {\rm{Pu}}$$^*$ (right).
The histograms are the results of present 4D Langevin calculation for excitation energy of 7 MeV, while the open circles are evaluated values of independent fission yields, namely post-neutron yields, given in JENDL/FPY-2011\cite{JENDLFPY2011} for thermal-neutron induced reactions on $^{235}$U and $^{239}$Pu.
In addition,
 the $+$ symbols are values given by Okumura {\it et al.}\cite{Okumura} and Schillebeeckx {\it et al.}\cite{Schillebeeckx240PuFPY} for thermal-neutron induced pre-neutron emission
 yields on $^{235}$U and $^{239}$Pu, respectively.
}
\label{FFMDs}
\end{figure}

Fig.~\ref{FFMDs} compares the calculated FFMDs for $^{236}$U$^*$ and $^{240}$Pu$^*$ with those for n + $^{235}$U and n + $^{239}$Pu reactions for which independent fission yield data were taken from JENDL/FPY2011\cite{JENDLFPY2011}, and 
those for pre-neutron emission were taken from
 Okumura {\it et al.}\cite{Okumura} and Schillebeeckx {\it et al.}\cite{Schillebeeckx240PuFPY}. 
The present calculations reproduce the width and the peak positions, and peak-to-valley ratio, reasonably well, if not perfect. 
Note that the present results are given for pre-neutron emission, thus keep a perfect mirror symmetry around $A_c/2$,
while data from JENDL/FPY2011 are for post-neutron emission. We can adjust model parameters to make agreement of the calculation with experimental data better, but it must be done not only looking at the FFMD but also for other quantities including TKE.  We will leave that task for our future work.

\begin{figure}[htbp]
\centering\includegraphics[scale=0.55,angle=270]{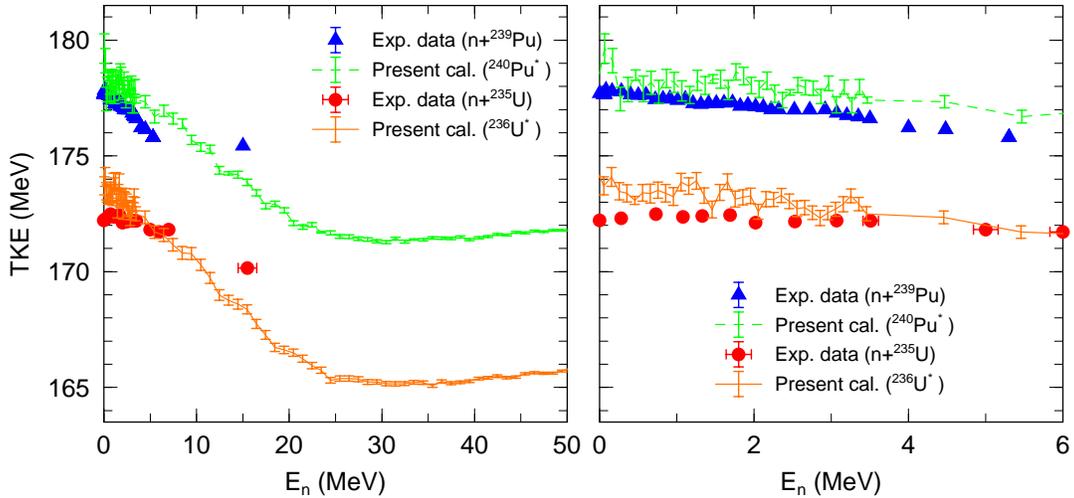}
%
\caption{(Color online) Average TKE (Total Kinetic Energy) of fission fragments for fissioning systems of $^{236} {\rm{U}}$$^*$ 
and $^{240} {\rm{Pu}}$$^*$. 
Two panels are the overall (the left) and enlarged (the right) views.
The scale of the vertical axis is common in these panels.
The solid lines with error bars denote our calculation, while symbols are experimental data~\cite{Dyachenko,Surin,Akimov}.
Note that we omit the multichance fission effects while the experimental 
data above several 
MeV include its effects.
}
\label{aveTKE}
\end{figure}

Fig.~\ref{aveTKE} exhibits the average TKE of fission fragments for $^{236}$U$^*$ and $^{240}$Pu$^*$ as a function of neutron energy for n + $^{235}$U and $^{239}$Pu reactions. The left and right panels in Fig.~\ref{aveTKE} correspond to the average TKE plot for neutron energy $E_n$ below 50 MeV and the enlarged view for $E_n \leq 6$ MeV, respectively.
Experimental data~\cite{Dyachenko,Surin,Akimov} are shown with filled triangles ($^{240}$Pu$^*$) and filled circles ($^{236}$U$^*$).  Results of the present calculation are given by a line with error bars, which indicates the statistical fluctuation in Monte-Carlo runs. 
As we can see, both of the calculated and measured average TKE show  clear decreasing trends as incident neutron energy (or excitation energy of compound nuclei) increases. 
This phenomenon has been recognized in experiments for a long time, but, at first glance, this is a trend against our intuition.
If the neutron energy increases, the excitation energy should increase. 
Then, it is natural to expect that energy to be shared to the kinetic energy of fission fragments should increase, which makes the average TKE increase.
But the experimental data show the opposite tendency, and our calculation can reproduce such behaviour in a quite reasonable manner.
It is the purpose of this paper to elucidate the reason for this decreasing trend of the average TKE of fission fragments as the neutron energy (and hence the excitation energy) increases. 
Furthermore, we can recognize that the calculated average TKEs agree well with the experiments within 1 MeV for neutron energies below several MeV, which indicates that the agreement is  accurate to the order of about 0.6\%, which is quite a good accuracy. 
For higher neutron energies, however, our calculation underestimates the measured TKE values. 
This disagreement is due to the fact that these data at the higher neutron energies include effects of the multichance fission, which makes the TKE values larger since it involves fission events for lower excitation energies. 
Inclusion of the multichance fission on average TKE is certainly an important issue, but we did not perform such a correction since we wish to comprehend in this work the reason why the average TKE of fission fragments decreases as the neutron energy increases without the complexity of multichance fission which may dilute the essence of this phenomena.  

\begin{figure}[htbp]
\centering\includegraphics[scale=0.75]{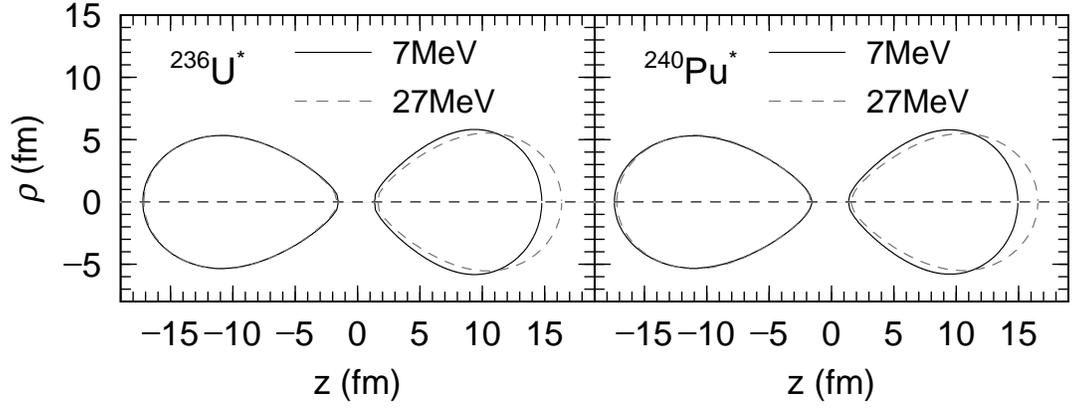}
\caption{(Color online) Shapes of fission fragments of $^{236} {\rm{U}}$$^*$ and $^{240} {\rm{Pu}}$$^*$
just afther scission for excitation energies of 7 MeV (solid lines) and 27 MeV (dash lines).
}
\label{shapes}
\end{figure}

Fig.~\ref{shapes} shows average shapes of fission fragments just after scission for $^{236} {\rm{U}}$$^*$ (left panel) and $^{240} {\rm{Pu}}$$^*$ (right panel) for excitation energies of 7 MeV (solid line) and 27 MeV (broken line). 
These shapes were obtained by averaging each of the four shape parameters at scission configuration, which is defined as neck radius, $r_{neck}$, $=$0, adjusting so that the heavy fragment  always stays at the right-hand side.
In each panel, the left and right fragments correspond to light and heavy fragments, respectively.
The volume of each fragment can be obtained by the integral $\pi\int\rho(z)^2dz$, where $\rho(z)$ denotes the value of $\rho$ coordinate corresponding to the nuclear surface as a function of $z$, see Fig.~\ref{shapes}.
The outer tip of the heavy fragment (right part) is closer to a sphere than the light fragment (left part), which is deformed well to prolate shapes for both excitation energies.
Moreover, it can be noticed that the light fragment shape does not depend on excitation energy noticeably. 
On the other hand, the heavy fragment shape depends on the excitation energy. When the excitation energy becomes larger, their shape deforms more to the prolate shape significantly. 
This fact is common for both $^{236}$U$^*$ and $^{240}$Pu$^*$ systems. 

\begin{figure}[htbp] 
\centering\includegraphics[scale=0.55,angle=270]{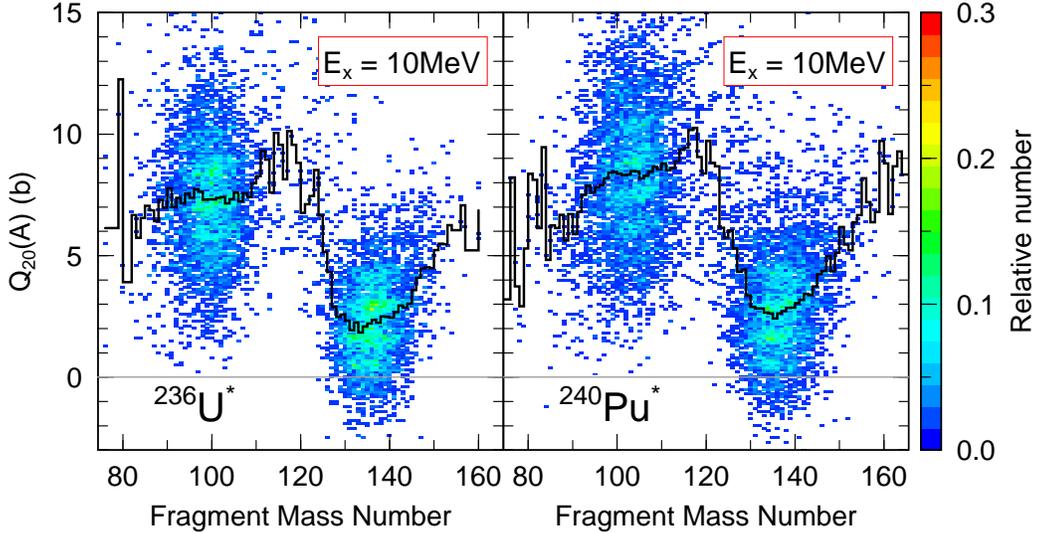}
\caption{(Color online) 
Quadrupole moment ($Q_{20}$) of fission fragments just after scission for the $^{236} {\rm{U}}$$^*$ (left panel) and $^{240} {\rm{Pu}}$$^*$ (right panel) fissioning systems at excitation energy of 10 MeV.
In each panel, the solid histogram shows average of $Q_{20}$ for each fission fragment mass number, namely, $\langle Q_{20}(A) \rangle$.
The scatter plot shows relative occurrence frequency of $Q_{20}$ by each fission event.
}
\label{q20}
\end{figure}

The scatter plot in fig.~\ref{q20} shows the distribution of the quadrupole moment $Q_{20}$ of the fission fragments originated from the compound nuclei $^{236}$U$^*$ and $^{240}$Pu$^*$, at an excitation energy of 10 MeV, as a function of a fragment mass number. 
In each panel, the averaged $\langle Q_{20} (A) \rangle$ is shown by a solid histogram.
The distribution of the $Q_{20}$ of the heavy fragments (A=130-144) is located at around 0 to 5 barns with $\langle Q_{20}(A) \rangle \simeq 2-3$ barns, while that of their counterpart formed as light fragments (A=90-110) is located at around 4 to 10 barns with $\langle Q_{20}(A) \rangle \simeq 7-8$ barns.
It indicates that the heavy fragments favor shapes closer to a sphere due to a strong shell effect around A=132-144 (double shell closure of $^{132}$Sn and deformed shell closures around A=144). 
The $\langle Q_{20} (A) \rangle$ shows a saw-tooth behaviour similar to the multiplicity of prompt neutrons.
Such features are commonly observed in $^{236}$U$^*$ and $^{240}$Pu$^*$. 

\begin{figure}[htbp] 
\centering\includegraphics[scale=0.78]{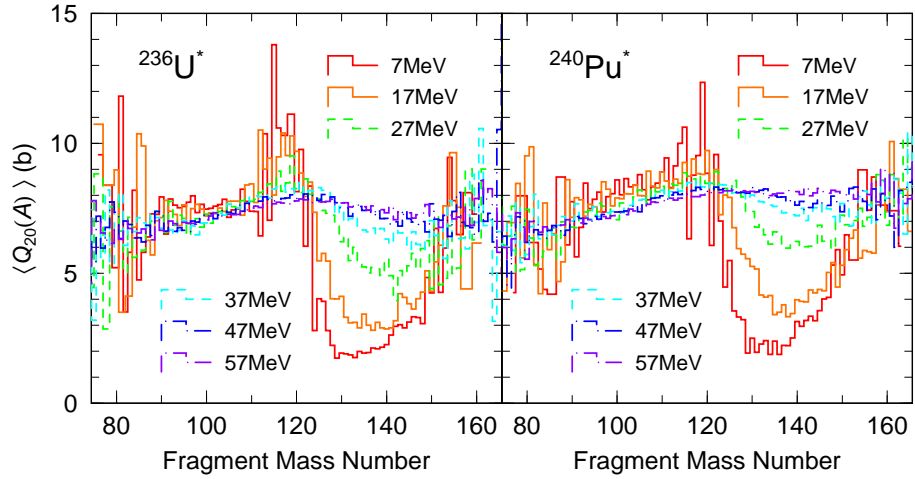}
\caption{(Color online) 
Dependence of the $\langle Q_{20}(A) \rangle$ of fission fragments 
originated from $^{236} {\rm{U}}$$^*$ (left panel) and $^{240} {\rm{Pu}}$$^*$ (right panel) for excitation energies $E_x = 7 - 57$ MeV. Note that the above results reflect only first chance fission.}
\label{q20-En}
\end{figure}

Fig.~\ref{q20-En} displays how this saw-tooth structure of $\langle Q_{20}(A) \rangle$ depends on excitation energy, starting from 7 MeV to 57 MeV. 
We notice that, besides large fluctuation at low excitation energies due to low Monte-Carlo statistics, the $\langle Q_{20}(A) \rangle$ of lighter fragments ($A \lesssim 126$) are almost independent of the excitation energy, while those of heavier fragments ($A \gtrsim 126$) clearly depend on it. 
The $\langle Q_{20}(A) \rangle$ of heavy fragments exhibits
a deep dip at around A=130 at an excitation energy of 7 to 17 MeV which is obviously caused by the spherical magicity of $^{132}$Sn, but this dip disappears gradually as excitation energy gets increased. 
Therefore, the dip, brought by the shell effect, gradually vanishes, since the heavy fragments can possess more energy as a form of "deformation energy" as excitation energy of the whole system is increased. 
This gradual change of the quadrupole moment of the heavier fragment is a manifestation of the washing out of the shell effects in nuclear fission. 
At excitation energies above 47 MeV, the shape of the heavy as well as light fragments do not change anymore, indicating that
the fission mechanisms are entering into the domain which can be described by the liquid-drop model. 
In this domain, both the light and heavy fragments are deformed to certain equilibrium prolate shapes.

\begin{figure}[htbp]
\centering\includegraphics[scale=0.78]{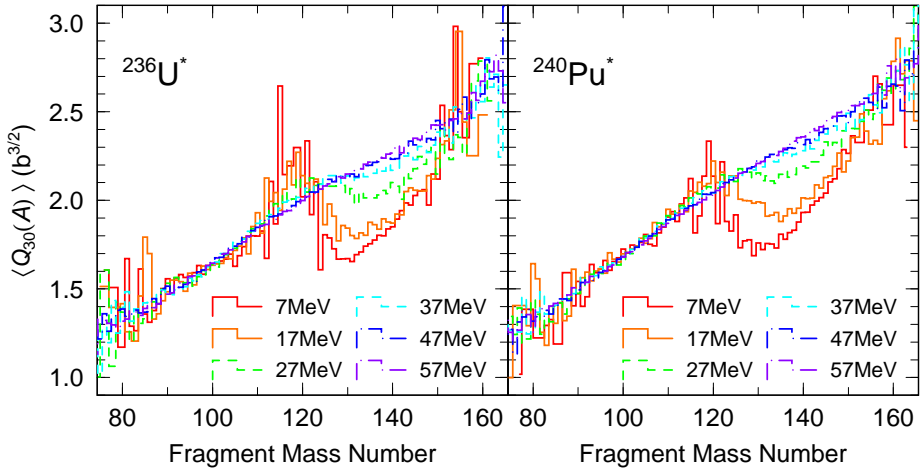}
\caption{(Color online) 
Dependence of the average octupole moment $\langle Q_{30}(A) \rangle$  of the fission fragments from $^{236} {\rm{U}}$$^*$ (left panel) and $^{240} {\rm{Pu}}$$^*$ (right panel) for excitation energies at $E_x = 7 - 57$ MeV. Note that the above results reflect only first chance fission.
}
\label{q30}
\end{figure}

Fig.~\ref{q30} shows a similar dependence as shown in fig.~\ref{q20-En} but for the octupole moment $\langle Q_{30}(A) \rangle$. 
In Fig.\ref{q30}, we found that the fission fragments have a noticeable octupole deformation (pear-like shapes). This result is in good agreement with the microscopic study~\cite{Scamps} where the importance of octupole deformation was discussed.
The excitation-energy dependence of $\langle Q_{30}(A) \rangle $
is very similar to that of $\langle Q_{20}(A) \rangle$, though $\langle Q_{30}(A) \rangle$ increases for a larger fragment mass number more than $\langle Q_{20}(A) \rangle$ , and saw-tooth structure is less pronounced for $\langle Q_{30}(A) \rangle$.
The $\langle Q_{30}(A) \rangle$ of lighter fragments is almost independent of the excitation energy of a compound nuclei, while that of the heavier fragments has a dip at low excitation energy but this dip is washed out as excitation energy gets larger.
At excitation energy above 47 MeV, the  $\langle Q_{30}(A) \rangle$ increases almost linearly as a function of the mass number of the fission fragments.

\begin{figure}[htbp]
\centering\includegraphics[scale=0.6]{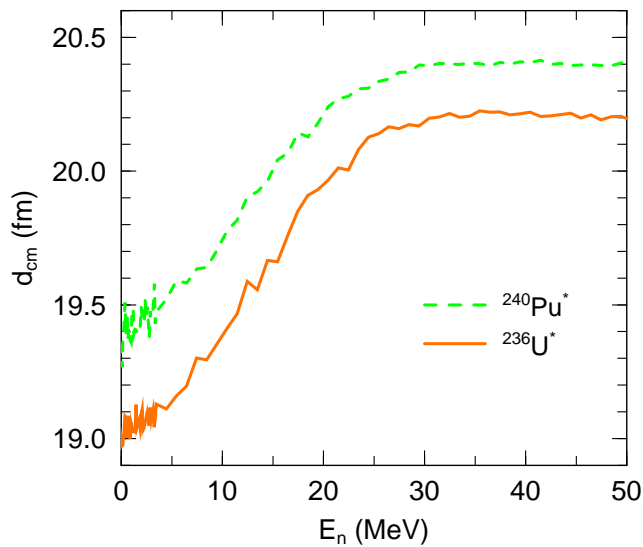}
\caption{(Color online) 
Distance between the center of mass $d_{cm}$ of the nascent fragments just after scission for $^{236} {\rm{U}}$$^*$ (solid line) and $^{240} {\rm{Pu}}$$^*$ (broken line) fissioning systems. Note that the above results reflect only first chance fission.
}
\label{dcm}
\end{figure}

\begin{figure}[htbp]
\centering\includegraphics[scale=0.35]{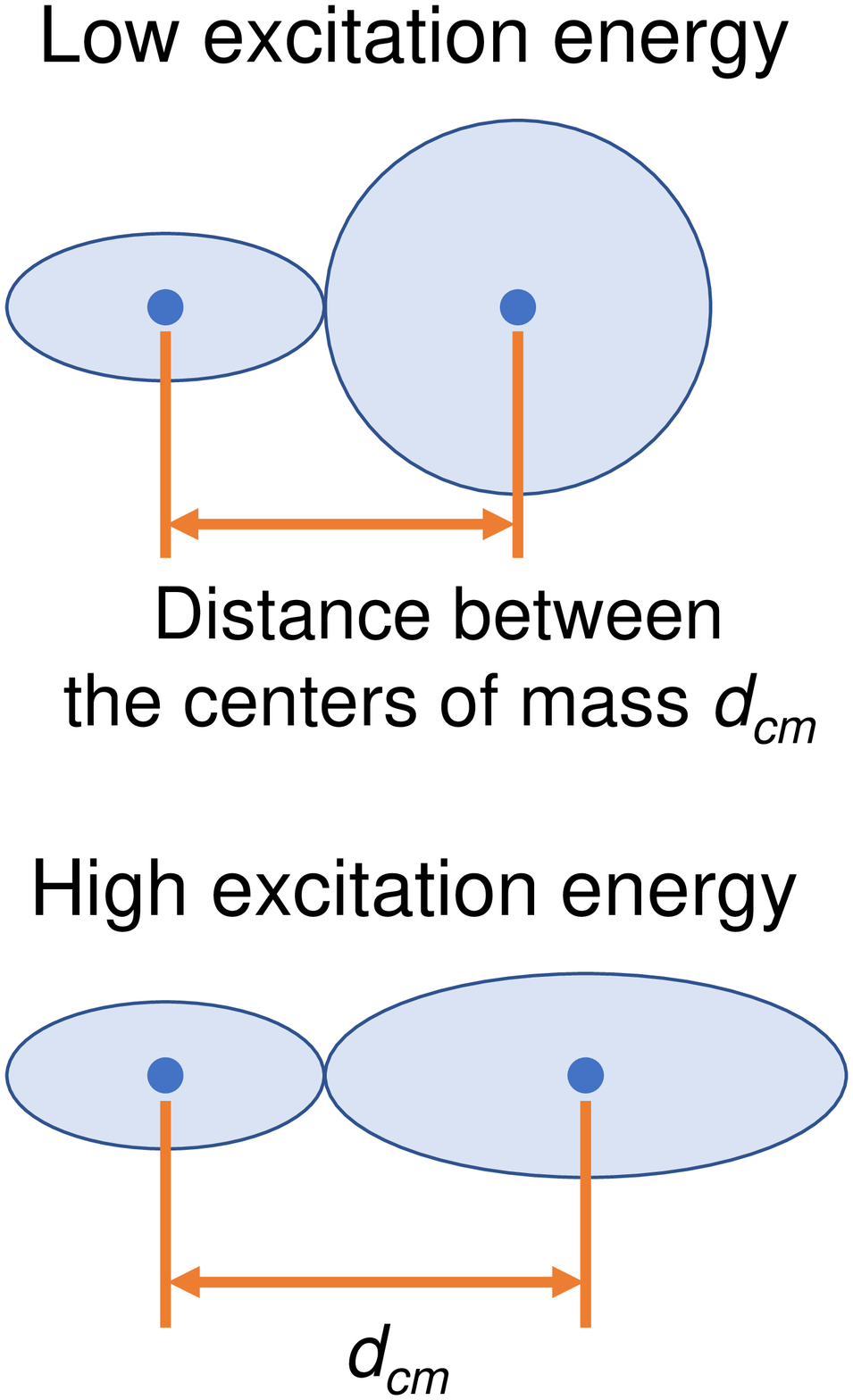}
\caption{(Color online) 
Schematic diagrams of shape of fission fragments just after scission for low excitation energy (upper) and high excitation energy (lower).
In each case, the shape of the light fission nucleus (on the left) is prolate.
On the other hand, the heavy fission fragments (on the right) is nearly sphere at low excitation energy, which turns out to be prolate at high excitation energy.
Therefore, the distance between the centers of mass $d_{cm}$ between nascent fragments just after scission is larger at high excitation energy than at low excitation energy, causing average TKE to decrease to higher excitation energy. 
}
\label{dcm-img}
\end{figure}

Based on the above results, we elucidate the reason why the average TKE of fission fragments decreases as the excitation energy increases. 
Fig.~\ref{dcm} shows the averaged distance between centers of mass of 2 nascent fragments just after the scission, $d_{\rm cm}$. 
Fluctuation of these lines from smooth ones is brought about by the Monte-Carlo statistics.
Thus only the smooth trend is of our interest.
We can observe that $d_{\rm cm}$ increases as the excitation energy rises in both compound nuclei $^{236}$U$^*$ and $^{240}$Pu$^*$. 
This increase is brought about by changing the shape of the heavier fragments from the near sphere for low excitation energies to a well deformed prolate shape at higher excitation energies.
This situation is schematically illustrated in fig.~\ref{dcm-img}. 
Low excitation energy favors the pair of nearly spherical heavy fragments and deformed light fragments.
When the excitation energy rises, the heavy fragment will be deformed.
Therefore, the distance between centers of mass of two fragments becomes larger for higher excitation energy. 
Such an increase in $d_{\rm cm}$ causes a decrease in TKE at higher excitation/neutron energies, because the Coulomb repulsion energy between two nascent fragments, which is the main source of the TKE of fragments, as calculated by
\begin{eqnarray}
TKE_{Coulomb} &=& 
{\frac{Z_L Z_H e^2}{d_{cm}}}  \ , 
\end{eqnarray}
should decrease as $d_{cm}$ gets increased, where $Z_L$ and $Z_H$ are proton numbers of the light and heavy fragments.  
Then, we can conclude that the reduction of the average TKE toward higher excitation energies is caused, at least as an important reason, by the washing out of the shell effect affecting shape of the heavy fragments to be nearly spherical at low excitation energy.

\section{Summary}
We have studied the reason why the average total kinetic energy of fission fragments decreases as the excitation energy of the compound nuclei increases, as indicated by experimental data of neutron-induced fission.
As the typical examples, we take $^{236}$U$^*$ and $^{240}$Pu$^*$ fissioning systems, or neutron-induced fission of $^{235}$U and $^{239}$Pu nuclei. 
For this sake, we have used a computational method based on the four-dimensional Langevin equations of nuclear fission~\cite{Ishizuka2017}.
In this method, Langevin equations for four dynamical variables representing nuclear shape during the scission process are solved in a Monte Carlo manner.
The free energy, inertia, and friction tensors to be used as transport coefficients in the Langevin equations were calculated by the two-center Woods-Saxon model, Werner-Wheeler approximation and Wall-and-Window formula, respectively.
To get a quantitative agreement of the absolute value of the TKE values with experimental data, the dumping factor $k_s$ for the Wall-and-Window friction was adjusted to $k_s$=0.55, and we have introduced different zero-point energy for each variable, as was originally pointed out by pioneering works by Bohr-Wheeler\cite{Bohr-Wheeler} and Hill-Wheeler\cite{Hill-Wheeler}.
This prescription does not change the dependence of the TKE on excitation energy, therefore the conclusion of this work.

We have found that calculated values of the average TKE of fission fragments indeed decrease as the incident neutron energy increases, which is in good accord with the trends in experimental data.  
Especially, the calculated TKE values agree with the data within 1 MeV for neutron energy below 6 MeV, which is in 0.6\% accuracy.
However, the calculated TKEs underestimate the experimental data existing at around $E_n=16$ MeV. 
A reason for this discrepancy is definitely due to the presence of multichance fission in the experimental data, which was ignored in the present calculation.

Then, we have elucidated the reason why the average TKE of fission fragments to decrease as the incident neutron energy (or excitation energy of the compound nuclei) increases. 
For this sake, we have calculated the quadrupole moment $Q_{20}$ of fission fragments just after scission.  We have found that $Q_{20}$'s of light and heavy fragments are distributed in a different manner; those for the light fragments are larger on the average than those for the heavy fragments, showing that the shape of the heavy fragments is affected by the magicity of the spherical $^{132}$Sn and 
deformed one around A=144.  We have also obtained $\langle Q_{20}(A) \rangle$, which shows average of the $Q_{20}$ for each fragment mass number $A$ and found that this quantity shows a clear saw-tooth structure similar to that of the multiplicity of prompt neutrons for low excitation energy.  However, a valley around $A=130$ in this saw-tooth structure is washed out as the excitation energy gets larger, and almost disappears for excitation energy larger than 30 MeV.  It indicates that the shape of the heavy fragments changes from nearly spherical shape at low excitation energy to a
well deformed prolate shape for higher excitation energy. This causes distance of the center-of-mass of the nascent fragments just after scission to increase as the incident neutron energy, thus the excitation energy of the compound nuclei, increases.  This causes the Coulomb repulsion between the nascent fragments to decrease, which results in the decrease of the average TKE of the fission fragments toward higher excitation energy.  

In the current study, we did not take account of the effects of multichance fission. Inclusion of the multichance fisssion must make the dependence of the average TKE on the incident neutron energy smaller.  Correction to include this effect is ongoing, and it will be discussed in detail in the next publication.  Furthermore, agreement of the calculated and evaluated mass distribution of fission fragments remained to be qualitative rather than quantitative in this work.  We are aware of the fact that this quantity is sensitive to combination of model parameters.  We recognize that making agreement of the mass distribution with experimental data simultaneously with that of the TKE observables to a quantitative level is also an important future work, and
an effort in this direction is also already ongoing.

\section{acknowledgement}
The authors are grateful to Prof.~J.~Maruhn of Frankfurt university for an essential support to carry out this study.
This study is supported by Japan Society for the Promotion of Science (JSPS) KAKENHI Grant No. 18K03642 and 21H01856.

\end{document}